\documentclass[11pt,onecolumn]{article}
\usepackage{graphicx,amssymb,amsmath,amsthm,hyperref}
\usepackage{geometry}
\title{A Model for Calculating Cost of Applying Electronic Governance and Robotic Process Automation to a Distributed Management System}

\author{Bonny Banerjee\\Independent Scholar and Consultant, USA\\ bonnybanerjee@yahoo.com
\and
Saurabh Pahune\\Cardinal Health, Dublin OH 43017, USA\\ saurabh.pahune@cardinalhealth.com
}

\date{\today}

\begin{document}

\maketitle

\begin{abstract}
  \textit{Electronic Governance} (eGov) and \textit{Robotic Process Automation} (RPA) are two technological advancements that have the potential to revolutionize the way organizations manage their operations. When applied to \textit{Distributed Management} (DM), these technologies can further enhance organizational efficiency and effectiveness. In this brief article, we present a mathematical model for calculating the cost of accomplishing a task by applying eGov and RPA in a DM system. This model is one of the first of its kind, and is expected to spark further research on cost analysis for organizational efficiency given the unprecedented advancements in electronic and automation technologies.
\end{abstract}


\section{Introduction}
\textit{Distributed Management} (DM) \cite{brumback2003end,bolden2011distributed} is a broad concept that refers to a decentralized approach to managing organizational resources and operations, in which decision-making authority and responsibility are distributed among different individuals, teams, or units based on skills, knowledge, expertise, needs and goals. This approach contrasts with traditional top-down management structures, in which decision-making authority is concentrated at the top of the organization and flows down to lower levels. DM allows for greater flexibility, agility, and responsiveness to changes in the business environment, as decisions can be made more quickly and at a more local level. DM can be facilitated by various technologies which enable individuals and teams to interact and share information and resources from anywhere, at any time. However, DM also poses challenges, such as the need for effective communication, coordination, and alignment across different units and teams, as well as the need for clear guidelines and protocols for decision-making and accountability.

In the context of an organization (as opposed to a government), \textit{Electronic Governance} (eGov) \cite{heeks2001understanding,egov2} refers to the use of information and communication technologies (ICTs) to improve the delivery of services or manufacturing of products, and enhance the operational efficiency and transparency of the organization \cite{egov,le2020participation}. eGov can improve the flow of information between different levels of the organization, facilitate decision-making processes, and enhance organizational agility. For example, in the healthcare industry \cite{kovac2014health,dhatterwal2023robotic}, eGov can enable hospitals to better manage patient data \cite{muttoo2019changing,alotaibi2017impact}, coordinate care across different departments, and respond more quickly to changes in patient needs \cite{siau2009factors}. eGov presents ways to utilize ICTs to help organizations facilitate their daily administration and to provide better services or higher-quality products to customers. 

In the context of DM, eGov can be used to enable decentralized decision-making processes and improve communication and collaboration between teams and departments. One application of eGov in DM is the use of collaborative tools such as project management software, instant messaging platforms, and videoconferencing tools. These tools enable teams to communicate and collaborate more effectively, regardless of their physical location. For example, a study \cite{morrison2020challenges} found that the use of videoconferencing in distributed teams led to improved communication and collaboration, as well as higher levels of trust between team members.

\textit{Robotic Process Automation} (RPA) \cite{flechsig2022robotic,vijai2020future} is a software technology that enables organizations to automate repetitive, rule-based tasks, and processes \cite{hofmann2020robotic}. RPA has been shown to have a significant impact on reducing costs, improving accuracy, and increasing efficiency \cite{delloite}. For example, in the banking industry, RPA can automate loan processing, account reconciliation, and other back-office operations, freeing up employees to focus on higher-value tasks such as customer service and product development \cite{vanhanen2020automation,romao2019robotic}. A study \cite{flechsig2022robotic} found that the use of RPA in supply chain management can reduce costs and increase efficiency, while also improving customer satisfaction\cite{supply}. A study \cite{oktem2014usage} showed that the use of eGov in higher education can improve the efficiency of administrative processes, reduce costs, and enhance the quality of education. 

In the context of DM, RPA can be used to automate routine tasks and enable employees to focus on higher-value activities. One application of RPA in DM is the automation of administrative tasks such as data entry, invoice processing, and customer service inquiries. By automating these tasks, organizations can reduce costs and improve accuracy, as well as free up employees to focus on more strategic tasks \cite{doshi2020automated}. For example, a case study \cite{ibm} found that the implementation of RPA in a customer service center led to a 70\% reduction in processing time and a 50\% increase in productivity.

Thus, the combination of eGov and RPA can have a significant impact on organizational performance in DM. By leveraging eGov tools to enable decentralized decision-making processes and RPA to automate routine tasks, organizations can improve efficiency, reduce costs, and increase agility. However, the implementation of eGov and RPA in DM also faces a number of challenges that must be addressed in order to fully realize their potential. These include the need for clear communication and coordination processes, as well as concerns around data security and privacy. Additionally, the implementation of these technologies may require significant investment in terms of infrastructure, training, and change management. As the technologies for eGov and RPA continue to evolve, it is likely that new applications and innovations will emerge, and that the impact of these technologies on organizational performance in DM will continue to be a topic of research and debate.

In summary, the application of eGov and RPA to DM has immense potential and some challenges. This calls for a formal model for cost analysis, which is missing. In this paper, we explore a mathematical model for calculating the cost of accomplishing a task by applying eGov and RPA in a DM system. This is the first model of its kind, and is expected to spark further research on cost analysis for organizational efficiency given the unprecedented advancements in electronic and automation technologies.

\section{Cost Calculation Model}

In \cite{gibson2015mathematical}, the authors investigate a mathematical model to generate theoretically consistent relationships between economic performance and organizational scale and structure. We will follow that model of DM in this paper. 

Let the depth $D$ be the number of levels of management in an organization. Level 1 denotes the highest level (e.g., CEO). $D$ is also the number of levels in the hierarchy below the top level as there is a bottom layer of employees without management duties. Employees in this bottom level are organized in work groups, each group reporting to a manager. For example, an organization of depth $D=5$ might consist of CEO, vice presidents, upper managers, middle managers, line managers, and line workers, stated in order from level $1$ (CEO) through $5$ (line managers); line workers do not have management duties.

Let $sl$ be a measure of employee skill level. The wage and benefit cost per unit time (e.g., per hour) of a line worker is

\begin{equation}
\label{Equ:Cost per unit time of a line worker}
C_{D+1} = Min\$ + \frac{1}{2} \triangle\$ \times D \times sl    
\end{equation}

\noindent where $Min\$$ and $\triangle\$$ are dollar values chosen such that Equ. \ref{Equ:Cost per unit time of a line worker} models how minimum employee cost varies as a function of the organization's depth and employee skill level. 

The wage and benefit cost per unit time of a manager at level $i$ ($i = 1, 2, ..., D$) is

\begin{equation}
\label{Equ:Cost per unit time of a manager}
C_i = C_{D+1} \times (1 + r)^{D-i+1} 
\end{equation}

\noindent where $r = r_o \times D$ ($r_o > 0$), implying that the ratio between salaries, $1+r$, at adjacent levels is an increasing function of overall organization depth. Then the cost per unit time of accomplishing a task is 

\begin{equation}
\label{Equ:Cost per unit time for a task by humans}
C^{(\text{humans})} = \frac{\text{Total cost}}{\text{Total time}} = \sum_{i=1}^{D+1}\sum_{j=1}^{e_i} C_{ij}\times t_{ij} \Bigg/ \sum_{i=1}^{D+1}\sum_{j=1}^{e_i} t_{ij} = \sum_{i=1}^{D+1}\sum_{j=1}^{e_i} C_{ij}\times p_{ij}
\end{equation}

\noindent where $e_i$ ($i=1, 2, ..., D$) is the number of managers at level $i$, $e_{D+1}$ is the number of line workers, $C_{ij}$ and $t_{ij}$ are respectively the cost per unit time of and the time contributed by the $j^{\text{th}}$ employee at level $i$, and $p_{ij} = t_{ij} / \sum_i \sum_j t_{ij}$ is the proportionate time contributed by that employee. Since $t_{ij}\geq 0$, $p_{ij}\geq 0$. Also, $\sum_i \sum_j p_{ij} = 1$.







For eGov, the cost of non-electronic processing includes all associated expenses such as labor, paper, printing, mailing, storage, and overhead. The cost of electronic processing includes expenses due to online forms, software, digital storage, and overhead. Similarly for RPA, the cost of manual processing includes expenses such as labor, training, human errors, and overhead. The cost of RPA includes the cost of purchasing, implementing, maintaining, and updating the RPA system.

We assume that artificial agents or bots work alongside human employees in an organization, and help to execute RPA and eGov. Just as a human employee, a bot has a certain skill level, works at a certain level in the organizational hierarchy, and incurs a certain cost per unit time. In order to calculate the cost of applying RPA and eGov to a DM system, we extend the above model to take into account the cost of human and bot employees.

We compute the cost per unit time of accomplishing a task by considering the time contribution of human employees and bot employees at a macro level. The cost per unit time of a bot line worker and a bot manager can be calculated in the same way as Equ. \ref{Equ:Cost per unit time of a line worker} and \ref{Equ:Cost per unit time of a manager} respectively:  

\begin{equation}
\label{Equ:Cost per unit time of a bot line worker}
C'_{D+1} = Min\$' + \frac{1}{2} \triangle\$' \times D \times sl'    
\end{equation}

\begin{equation}
\label{Equ:Cost per unit time of a bot manager}
C'_i = C'_{D+1} \times (1 + r')^{D-i+1} 
\end{equation}

\noindent where $'$ is used to distinguish the corresponding quantities between human and bot employees. Following Equ. \ref{Equ:Cost per unit time for a task by humans}, the cost per unit time of accomplishing a task by bot employees is

\begin{equation}
\label{Equ:Cost per unit time for a task by bots}
C^{(\text{bots})} = \sum_{i=1}^{D+1}\sum_{j=1}^{e'_i} C'_{ij}\times p'_{ij}
\end{equation}

Then the cost per unit time of accomplishing a task jointly by human and bot employees is

\begin{equation}
\label{Equ:Cost per unit time for a task by bots and humans}
\mathcal{C} = \kappa \times (C^{(\text{humans})} + C^{(\text{bots})})
\end{equation}

\noindent where $\kappa$ is a normalizing constant. Expanding Equ. \ref{Equ:Cost per unit time for a task by bots and humans}, we get 

\begin{equation}
\mathcal{C} = \kappa \times \Bigg(\sum_{i=1}^{D+1}\sum_{j=1}^{e_i} C_{ij} \times p_{ij} + \sum_{i=1}^{D+1}\sum_{k=1}^{e'_i} C'_{ik} \times p'_{ik}\Bigg) = \sum_{i=1}^{D+1} \Bigg(\sum_{j=1}^{e_i} C_{ij} \times q_{ij} + \sum_{k=1}^{e'_i} C'_{ik}\times q'_{ik}\Bigg)
\end{equation}

\noindent where $\kappa = 1/ \sum_i (\sum_j p_{ij} + \sum_k p'_{ik})$, $q_{ij} = \kappa\times p_{ij}$, and $q'_{ik} = \kappa\times p'_{ik}$. Therefore, $\kappa, q_{ij}, q'_{ik} \geq 0$, and $\sum_i (\sum_j q_{ij} + \sum_k q'_{ik}) = 1$. Interestingly, for line workers (bots and humans), i.e. when $i=D+1$, we have

\begin{align}
&\sum_{j=1}^{e_{D+1}} C_{D+1,j}\times q_{D+1,j} + \sum_{k=1}^{e'_{D+1}} C'_{D+1,k}\times q'_{D+1,k} \nonumber\\
&= \sum_{j=1}^{e_{D+1}} (Min\$ + \frac{1}{2} \times \triangle\$ \times sl_j) \times q_{D+1,j} + \sum_{k=1}^{e'_{D+1}} (Min\$' + \frac{1}{2} \times \triangle\$' \times sl'_k) \times q'_{D+1,k} \nonumber\\
&= Min\$ \times \sum_{j=1}^{e_{D+1}} q_{D+1,j} + \frac{1}{2} \times \triangle\$ \times \sum_{j=1}^{e_{D+1}}(sl_j \times q_{D+1,j}) \nonumber\\ 
&~~~~~~~~~~ + Min\$' \times \sum_{j=1}^{e'_{D+1}} q'_{D+1,j} + \frac{1}{2} \times \triangle\$' \times \sum_{k=1}^{e'_{D+1}}(sl'_k \times q'_{D+1,k}) \nonumber\\
&= \alpha + \beta \times \sum_{j=1}^{e_{D+1}}(sl_j \times q_{D+1,j}) + \alpha' + \beta' \times \sum_{k=1}^{e'_{D+1}}(sl'_k \times q'_{D+1,k})
\end{align}

\noindent That is, each line worker (human or bot) has a skill level, and the cost for accomplishing a task involves all skill levels in different time proportions.

\section{Conclusions}
The application of eGov and RPA to DM can improve organizational efficiency, productivity, and agility. However, it also faces a number of challenges that must be addressed in order to fully realize its potential. For example, it has been argued in \cite{bertot2008citizen} that successful implementation of eGov requires intricate planning and design, which may increase the cost of governance. The degree to which user satisfaction and cost reduction can be accomplished simultaneously through eGov is unclear. Thus, a formal model for cost analysis is necessary, but missing. We presented a mathematical model for calculating the cost of accomplishing a task by applying eGov and RPA in a DM system. As the areas of eGov, RPA and DM continue to evolve, it is likely that new applications and innovations will emerge, and their impact on organizational performance will continue to be a topic of research, which can be analyzed using our model.

\section{Authors' Contributions}
BB directed the research and derived the Cost Calculation Model (Section 2). SP reviewed the literature and wrote the Introduction (Section 1). Both authors contributed to the Abstract, Keywords, and Conclusions.

\bibliographystyle{ieeetr} 
  \bibliography{dms}
 
\end{document}